\input harvmac
\input psfig
\newcount\figno
\figno=0
\def\fig#1#2#3{
\par\begingroup\parindent=0pt\leftskip=1cm\rightskip=1cm\parindent=0pt
\global\advance\figno by 1
\midinsert
\epsfxsize=#3
\centerline{\epsfbox{#2}}
\vskip 12pt
{\bf Fig. \the\figno:} #1\par
\endinsert\endgroup\par
}
\def\figlabel#1{\xdef#1{\the\figno}}
\def\encadremath#1{\vbox{\hrule\hbox{\vrule\kern8pt\vbox{\kern8pt
\hbox{$\displaystyle #1$}\kern8pt}
\kern8pt\vrule}\hrule}}
\def\underarrow#1{\vbox{\ialign{##\crcr$\hfil\displaystyle
 {#1}\hfil$\crcr\noalign{\kern1pt\nointerlineskip}$\longrightarrow$\crcr}}}
%
\overfullrule=0pt

%

\def\S{{\bf S}}
\def\R{{\bf R}}

\font\zfont = cmss10 

\def\bigone{\hbox{1\kern -.23em {\rm l}}}
\def\ZZ{\hbox{\zfont Z\kern-.4emZ}}

\Title{hep-th/9807109, IASSNS-HEP-98-65}
{\vbox{\centerline{Theta Dependence in The Large $N$ Limit}
\bigskip
\centerline{Of Four-Dimensional Gauge Theories}}}
\smallskip
\centerline{Edward Witten}
\smallskip
\centerline{\it School of Natural Sciences, Institute for Advanced Study}
\centerline{\it Olden Lane, Princeton, NJ 08540, USA}\bigskip

\medskip

\noindent
The $\theta$  dependence of pure gauge theories in four dimensions can be 
studied
using a duality of large $N$ gauge theories with string theory on a certain
spacetime.  Via this duality, one can argue that for every $\theta$, there
are infinitely many vacua that are stable in the large $N$ limit.  The true
vacuum, found by minimizing the energy in this family, is a smooth function of
$\theta$ except at $\theta=\pi$ where it jumps.  This jump is associated
with spontaneous breaking of CP symmetry.  Domain walls separating adjacent
vacua are described in terms of wrapped sixbranes.  
\Date{July, 1998}

\newsec{The Problem}

In weak coupling, the dependence of four-dimensional gauge theories
on the theta angle is computed via instantons.  An instanton
or anti-instanton contribution is proportional to
$\exp\left(-8\pi^2/g^2\right)\exp(\pm i\theta)$.  For example, in
  spontaneously broken gauge
theories, in which instantons have a characteristic maximum size,
and a characteristic effective coupling, the $\theta$
 dependence is determined by an instanton expansion.

For unbroken asymptotically free
gauge theories, the situation is rather different.
In such theories, at the classical level, instantons come in all sizes.
The infrared behavior of the instanton gas is difficult to understand,
and it is not clear what effective description should be used at long
wavelengths to describe the theta dependence, or other aspects of the
physics.

One sharp way to pose the question is to consider the large $N$ limit
of an $SU(N)$ gauge theory \ref\thooft{G. 't Hooft, ``A Planar Diagram
Theory For Quark Confinement,'' Nucl. Phys. {\bf B72} (1974) 461.}.
The large $N$ limit is an important avenue for understanding
the dynamics of pure gauge theory, or gauge theory
with a small number of light quark flavors, in four dimensions.  The large $N$
limit is attained by taking $N\to\infty$ with $\lambda=g^2N$ fixed.
Thus the amplitude for an instanton or anti-instanton of definite size
 is weighted by a factor
of $\exp(-8\pi^2N/\lambda)$, and it appears that instanton effects would
vanish exponentially for $N\to\infty$.  However, there are a variety
of reasons to believe that, because of infrared divergences,
 this is not so.\foot{In contrast, ${\cal N}=4$ super Yang-Mills
theory, which is scale-invariant rather than asymptotically free, does not
have these infrared divergences, and does have exponentially small 
$\theta$-dependent
effects, 
which can be computed via instantons \ref\colors{M. Bianchi,
M. B. Green, S. Kovacs, and G. Rossi, ``Instantons In Supersymmetric
Yang-Mills and $D$-Instantons In IIB Superstring Theory,'' hep-th/9807033.}.
These instantons are related in the AdS/CFT correspondence to $-1$-branes
\ref\gb{T. Banks and M. B. Green, ``Nonperturbative Effects in
$AdS_5\times S^5$ String Theory And $d=4$ SUSY Yang-Mills,''
hep-th/9804170.}.}  For example,
if light quarks are included,
then the $\theta$ dependence can be seen in current algebra
\nref\baluni{V. Baluni, ``CP Violating Effects In QCD,'' Phys. Rev.
{\bf D19} (1979) 2227.}
\nref\witten{E. Witten, ``Large $N$ Chiral Dynamics,''  Annals of Physics
{\bf 128} (1980) 363}
\refs{\baluni,\witten}, by reinterpreting some old pre-QCD computations
\nref\dashen{R. Dashen, Phys. Rev. {\bf D3} (1971) 1879.}
\refs{\dashen}. One can show that if chiral symmetry breaking
survives in the large $N$ limit, then so does the  theta dependence.
Moreover, the most plausible interpretation of how the solution of the
$U(1)$ problem of QCD fits into the $1/N$ expansion implies that
in the pure  gauge theory, the $\theta$ dependence of the ground state
energy is present to leading order\foot{An explanation of exactly
what is meant by ``leading order'' will emerge  below.}
 in  $1/N$ 
\ref\wittentwo{E. Witten, ``Current Algebra Theorems For The $U(1)$
`Goldstone Boson,' '' Nucl. Phys. {\bf B156} (1979) 269.}.
These arguments suggest that, as in some two-dimensional models where
somewhat similar questions can be asked
\nref\div{A. D'Adda, M. Luscher, and P. DiVecchia, Nucl. Phys. {\bf B146} 
(1979) 
63.}
\nref\wittenthree{E. Witten, Nucl. Phys. {\bf B149} (1979) 285.}
\refs{\div,\wittenthree}, the $\theta$ dependence of pure gauge theory
in four dimensions
(with or without a small number of matter fields in the fundamental
representation of the gauge group)
is present in the leading order of the $1/N$ expansion. 

If so, one can draw an interesting deduction about the form of the
theta-dependence \witten.  In any theory with $N\times N$ matrix fields
$\Phi_i$, the large $N$ limit is obtained by taking a Lagrangian of the
form
\eqn\relag{L(\Phi_i)=NS(\Phi_i;w_\alpha),}
where $S$ is independent of $N$ and the 
$w_\alpha$ are parameters such as bare masses and coupling constants.
With the normalization in \relag, the large $N$ limit is obtained
by keeping $w_\alpha$ fixed as $N\to \infty$.

Now instead, the most general renormalizable Lagrangian for gauge
fields in four dimensions takes the form
\eqn\elag{L={N\over 4\lambda}\Tr F_{\mu\nu}F^{\mu\nu}
+{\theta\over 16\pi^2}\epsilon^{\mu\nu\alpha\beta}\Tr F_{\mu\nu}F_{\alpha
\beta}.}
The normalization is chosen so that $\theta$ is an angular variable.
The general recipe of \relag\ would tell us to set $\theta=N\psi$
and keep $\psi$ fixed for $N\to\infty$.  In the large $N$ limit,
the vacuum energy $E$ is proportional to $N^2$ (as the number of degrees of
freedom is of that order).  So we expect $E(\psi)=N^2h(\psi)$ for
some function $h$ which should have a limit as $N\to\infty$.
  In terms of $\theta$, that means
\eqn\hudc{E(\theta)=N^2h(\theta/N).}
In addition, $E$ must obey
\eqn\udc{E(\theta)=E(\theta+2\pi).}
These conditions are, however, practically incompatible: a smooth
function of
$\theta/N$ cannot be invariant to $\theta\to \theta+2\pi$ unless it is 
constant.

The most plausible way out seems to be \witten\
that $E(\theta)$ is a multibranched function because
of many candidate vacuum states that all become stable (but not degenerate)
for $N=\infty$.  
Such behavior occurs in many two-dimensional models
\ref\coleman{S. Coleman, ``More On The Massive Schwinger Model,''
Ann. Phys. (NY) {\bf 101} (1976) 239.} (including \wittenthree\ 
some with $1/N$ expansions that raise
issues like those we are discussing here).
In the $k^{th}$ vacuum, the energy would be
\eqn\upud{E_k(\theta)=N^2h((\theta+2\pi k)/N).}
The truly stable vacuum would be found, for each $\theta$, by minimizing
$E_k$ with respect to $k$.  The actual vacuum energy would be therefore
\eqn\hicco{E(\theta)=N^2\min_{k}h((\theta+2\pi k)/N).}
This function is periodic in $\theta$, but (if $h$ is not constant) it
is not smooth -- at some value of $\theta$ there is a jump between
two different ``branches.''

Under a CP transformation,
one has $\theta\to -\theta$.  So in particular CP is a symmetry
if and only if $\theta$ equals 0 or $\pi$.    Hence
$h(\theta)=h(-\theta)$.  Note that CP acts by $k\to -k$ at
$\theta=0$, and by $k\to -1-k$ at $\theta=\pi$.

Moreover, $E(\theta)$ has its absolute
minimum at $\theta=0$, because precisely at $\theta=0$ the integrand
of the Euclidean space path integral is real and positive.\foot{
At $\theta=0$
all contributions to the path integral receive positive weights; at 
$\theta\not=0$, this is not so because  the instanton factor is  $e^{i\theta}$.
The Euclidean space path integral in volume $V$ computes $\exp(-VE(\theta))$,
so $E(\theta)$ is minimized by maximizing the Euclidean space path
maximal, which happens when the weights are all positive.}
If the vacuum is unique at $\theta=0$, then the minimum in \hicco\ occurs
for $k=0$ (otherwise $k$ and $-k$ would both contribute).
Moreover, one expects that $d^2h/d\theta^2\not= 0$ at $\theta=0$
because of arguments involving the $U(1)$ problem in the theory with
quarks
\wittentwo, or simply because of the absence of a symmetry
that would make this quantity vanish.
If so, we can set $h(\theta)=C\theta^2+\dots$, where $C$ is positive
and the higher order terms do not contribute to \hicco\
in leading order in $1/N$.
Thus, one conjectures that the large $N$ structure of the vacuum
energy is
\eqn\doofo{E(\theta)=C\min_k\left(\theta+2\pi k\right)^2+O(1/N),}
with some constant $C$.
This function exhibits a nonanalyticity at $\theta=\pi$, which we associate
with a jump between two vacua (with $k=0$ and $k=-1$)
and the spontaneous breaking of CP invariance.

\newsec{The Computation}

In what dynamical approximation to gauge theory
can one hope to check 
the ideas that were just reviewed?  All known approaches to the dynamics
of four-dimensional gauge theory in which one can exhibit any of the
difficult properties like confinement and the mass gap
involve replacing the theory by
a simpler theory which is hoped to be in the same universality class.
There are several candidates for what the simpler theory can be.
One candidate is  lattice gauge theory.  This framework makes
it possible to compute a great deal, especially upon using computer
simulation to get away from the strong coupling limit.  But it is
not very convenient for discussing the $\theta$ dependence,
particularly if one wishes to probe issues in which the invariance under
$\theta\to \theta+2\pi$ is important.  Another possibility is to consider
the realization of four-dimensional gauge theory via $M$-theory fivebranes
\ref\wittenfour{E. Witten, ``Solutions Of Four-Dimensional Field Theories
Via $M$-Theory,'' Nucl. Phys. {\bf B500} (1997) 3.}, where again, at the cost 
of replacing
the theory of interest by a simplified version that is hopefully in the same
universality class, one can demonstrate the mass gap and confinement.
We will instead study the problem in yet a third
framework in which one can exhibit the mass gap and confinement in the context
of a simplified version of four-dimensional gauge theory. 
\nref\malda{J. Maldacena, ``The Large $N$ Limit Of Superconformal Field
Theories And Supergravity,'' hep-th/9711200.}
\nref\kleb{S. Gubser, I. R. Klebanov, and A. M. Polyakov, ``Gauge Theory
Correlators From Noncritical String Theory,'' hep-th/9802109.}
\nref\wittenfive{E. Witten, ``Anti-de Sitter Space And Holography,''
hep-th/9802150.} 
This involves a circle of ideas connected with the
 correspondence between conformal field theory and
quantum gravity on Anti-de Sitter space \refs{\malda-\wittenfive}.

\def\R{{\bf R}}
\def\S{{\bf S}}
To get to the specific issues of interest here as quickly as possible,
we will begin as in section 4 of
\ref\wittensix{E. Witten, ``Anti-de Sitter Space,
Thermal Phase Transition, And Confinement In Gauge Theories,''
hep-th/9803131.} with Type IIA superstring theory on $M= \R^4\times \S^1\times
\R^5$, with $N$ parallel fourbranes whose worldvolume is  $V=
\R^4\times\S^1\times
x$; here $x$ is a point in $\R^5$.  We assume that the ``spin structure'' is
such that fermions change sign in going around the $\S^1$.  Then the
theory on the branes is at low energies a pure $U(N)$ gauge theory
in four dimensions.  It follows from the general AdS/CFT correspondence
that the large $N$ behavior of the $SU(N)$ part of this gauge theory can be
studied by studying weakly coupled
string theory on the supergravity solution $X$ which these branes generate.
Topologically $X=\R^4\times
D\times \S^4$, where $D$ is a two-dimensional disc.  
The change in topology from $M$ to $X$ is crucial (along with the fact
that $X$ is a smooth manifold without branes), both in the explanations
of confinement and the mass gap in \wittensix\ and in the discussion below
of the $\theta$ dependence.
The  metric of $X$ is
\eqn\dolfo{ds^2={8\pi\over 3}\eta\lambda^3\sum_{i=1}^4(dx^i)^2
+{8\over 27}\eta\lambda\pi\left(\lambda^2-{1\over \lambda^4}\right)d\psi^2
+{8\pi\over 3}\eta\lambda {d\lambda^2\over \lambda^2-{1\over \lambda^4}}
+{2\pi\over 3}\eta\lambda d\Omega_4^2.}
Here $x^i$ are coordinates on $\R^4$,
$\lambda$ and $\psi$ (with $1\leq\lambda\leq \infty$, $0\leq\psi
\leq2\pi$) are polar coordinates on $D$ (note that $\lambda=1$ is the
origin of the polar coordinates, the ``center'' of $D$), 
and $d\Omega_4^2$ is the metric of
a unit four-sphere.  $\eta$ is a parameter which determines how far
one is from conventional four-dimensional gauge theory; for $\eta>>1$,
the string theory on $X$ can be studied via long wavelength supergravity,
while asymptotically-free gauge theory is expected to emerge in the opposite
limit $\eta\to 0$.
In the present paper, we work in an approximation of long wavelength 
supergravity, so to compare with gauge theory,
we must assume that the system has no phase transition
as a function of $\eta$.

How can we include $\theta$ in the formulation of gauge theory via
fourbranes?  This can be done quite simply by including the $U(1)$
gauge field that arises in Type IIA superstring theory from the Ramond-Ramond
sector.  We will in this paper denote that field as $a$, and its
field strength as $f_{ij}=\partial_ia_j-\partial_ja_i$.  
Let us reconsider Type IIA superstring theory on $M=\R^4\times \S^1\times \R^5$
with the wrapped fourbranes of worldvolume $V$.  The
low energy worldvolume effective Lagrangian of the fourbranes has a term
\eqn\pooly{\Delta L = \int_V a\wedge {\Tr F\wedge F\over 8\pi^2}.}
(The most familiar manifestation of this term is that instantons
on the fourbrane are charged with respect to $a$ -- they carry zerobrane
charge.)  Here $F$ is the $U(N)$ field strength.  We now modify the Type IIA
vacuum so that $f=0$, but 
\eqn\humbly{\int_{\S^1}a=\theta_a} 
is possibly non-zero.  (The left hand side is gauge-invariant modulo
$2\pi {\bf Z}$, so we interpret $\theta_a$ as an angle.)
At low energies in four dimensions, \pooly\ reduces to a theta term
in the gauge theory action, and the four-dimensional Yang-Mills theta angle is
\eqn\joolop{\theta=\theta_a.}

What we have done so far is just to learn how to include $\theta$ in the
fourbrane description of four-dimensional gauge theory.
Now, we go over to the dual description by supergravity (or string
theory) on $X$.  
In doing so, we must bear in mind that the parameters of the theory
are determined by specifying the   Type IIA vacuum far away
from the branes, that is at large $\lambda$,
and then the behavior at small $\lambda$ is determined by the supergravity
equations on $X$
 (or, if $\eta$ is small, the full string theory on $X$)
and encodes the behavior of the gauge theory.  For example, in
the original description with branes on $M$ we assumed that $f=0$.
In the  dual description on $X$, the analogous statement
is that $f=0$
for  $\lambda\to\infty$.  Likewise,  \humbly\ should be interpreted to mean
that $\int_{\S^1}a=\theta_a=\theta$ if the integral is taken  at large 
$\lambda$.  If we combine these conditions (plus Stokes' theorem
$\int_Df=\lim_{\lambda\to\infty}\int_{\S^1}a$),\foot{
 $\int_Df$ is of course defined as $\int_Dd\lambda d\psi f_{\lambda
\psi}$.}  we learn that $\int_D f=
\theta_a=\theta$ mod $2\pi {\bf Z}$. The $2\pi {\bf Z}$ indeterminacy
arises because the left hand side is a well-defined real number,
but $\theta$ is an angle.  Hence 
\eqn\omicro{\int_D f = \theta+2\pi k}
for some integer $k$.

 Maxwell's  equations for the $f$ field have a normalizable
zero mode in which the only nonzero component is
\eqn\hjd{f_{\lambda\psi}={6\over \lambda^7}.}
The normalization has been chosen so that $\int_D f=2\pi$.  Hence, we can find
a solution of \omicro\ obeying Maxwell's equations in the simple form
\eqn\jicom{f_{\lambda\psi}=(\theta+2\pi k){3\over \pi\lambda^7}.}
The back reaction on the geometry produced by this $f$ field is negligible
in the limit that $N$ goes to infinity with fixed $\theta+2\pi k$.  The
reason for this is that the classical action of the spacetime is of order $N^2$
(like the vacuum energy of the large $N$ gauge theory to which it is dual).
The kinetic energy of the $f$ field is of relative order $1/N^2$ as it is 
simply 
\eqn\hormo{\int d^{10}x \sqrt g
f_{ij}f^{ij},} with no factors of $N$ or of the string coupling constant.
So to lowest order in $1/N$, we obey \omicro\ and the 
classical equations of motion
simply by solving for $f$ in the fixed spacetime $X$. 

The $\theta$ and $k$-dependent part of the vacuum energy is found by evaluating
\hormo\ with $f$ as given in \jicom.  It thus takes the
form foreseen in the introduction:
\eqn\jicn{E_k(\theta)=C(\theta+2\pi k)^2,}
with some positive constant $C$ that is independent of $N$.  The vacuum
energy for given $\theta$
is obtained by minimizing this with respect to  $k$, 
\eqn\picn{E(\theta)=C\min_k(\theta+2\pi k)^2.}

We have obtained precisely the structure anticipated in the introduction.
For given $\theta$, there are infinitely many vacua, labeled by the choice
of an integer $k$.  The vacuum energy $E(\theta)$ is a smooth function of
$\theta$ except at $\theta=\pi$, where there is a jumping between two
solutions ($k=0$ and $k=-1$).  This jumping represents spontaneous breaking
of CP at $\theta=\pi$.

Finally, we should check that the vacua labeled by $k$ are all stable
for $N\to\infty$, and find a framework for
 estimating their lifetime for finite $N$.
The essential point is to describe the domain wall separating vacua with
adjacent values of $k$.  
The decay of a $k$-vacuum  involves nucleation of
a ``bubble'' with a smaller  value of $|\theta+2\pi k|$ and hence
 a lower energy density; this bubble is bounded by a domain wall.
If the energy per unit area of such a domain wall
is large for $N\to\infty$, then the $k$-vacua have lifetimes that go to 
infinity
for $N\to\infty$.  

In fact, the domain wall is constructed simply  by compactifying a Type IIA
sixbrane on the $\S^4$ factor in $X=\R^4\times D\times \S^4$.  
Let $w$ be a point in $D$ and let
$C$ be a codimension one surface in $\R^4$ given by, say, $x^3=0$ 
(with $x^3$ one of the space coordinates).  Consider a sixbrane whose
worldvolume is $Q= C\times w\times \S^4$.  The value of $k$ jumps by $\pm 1$
in crossing such a sixbrane (the sign depends on orientations and on the 
direction
of crossing the sixbrane).  The defining property of a Type IIA sixbrane
is that if $E$ is a two-surface that has linking number 1 with the sixbrane
worldvolume, then $\int_E f=2\pi$.  In the present case, we can take 
$E=D_1-D_2$, where $D_1$ and $D_2$ are copies of $D$ that are respectively
to the ``left'' or ``right'' of $C$.\foot{In other words, $D_i=y_i\times
D\times z$, where $z$ is a point in $\S^4$, and the $y_i$ are points
in $\R^4$ to the left or right of $C$.}  So
\eqn\icox{\int_{D_1}f-\int_{D_2}f = \pm 2\pi,}
and this means that $k$ jumps by $\pm 1$ in crossing the sixbrane.  The domain
wall is thus not an ordinary soliton, as one might naively have thought, but
a $D$-brane.  In particular, color flux tubes associated with quark
confinement can terminate on the boundaries between different vacua,
just as they can terminate \wittenfour\ on chiral domain walls
in ${\cal N}=1$ supersymmetric gauge theory in four dimensions.  (Both
kinds of domain wall involve a $2\pi$ jump in $\theta$.)

Now, the energy density of a Type IIA sixbrane is for weak coupling
of order $1/{\lambda}_{st}$
($\lambda_{st}$ is the Type IIA string coupling constant), and as a result,
in the large $N$ limit it is of order $N$.  An instanton describing the
decay of a $k$-vacuum can be constructed as a sort of sixbrane bubble
and has an action that grows as a power of $N$. So the lifetime of a 
$k$-vacuum,
for any given $k$, is exponentially long for $N\to\infty$.

\bigskip
This work was supported in part by NSF Grant PHY-9513835.
I would like to thank S.-J. Rey for comments.
\listrefs
\end